\documentclass[twocolumn,showpacs,preprintnumbers,amssymb]{revtex4}
\usepackage{graphicx}
\usepackage{dcolumn}
\usepackage{bm}
\input epsf

\newcommand{\ovec}[1]{{\mbox{\boldmath $#1$}}}

\newcommand{\bA}{\ovec{A}}      

\newcommand{\bB}{\ovec{B}}

\newcommand{\be}{\ovec{e}}
\newcommand{\bE}{\ovec{\cal{E}}}

\newcommand{\bff}{\ovec{f}}

\newcommand{\bk}{\ovec{k}}

\newcommand{\bu}{\ovec{u}}

\newcommand{\bmB}{\overline{\ovec{B}}}

\begin{document}
\preprint{NORDITA 2002-47 AP}

\title{Contributions to the theory of a two--scale homogeneous dynamo experiment}
\author{Karl-Heinz R\"adler}
\affiliation{Astrophysical Institute of Potsdam,
An der Sternwarte 16, D-14482 Potsdam, Germany}
\author{Axel Brandenburg}
\affiliation{NORDITA, Blegdamsvej 17, DK-2100 Copenhagen \O, Denmark}
\date{\today,~ $ $Revision: 1.34 $ $}

\begin{abstract}
The principle of the Karlsruhe dynamo experiment is closely related 
to that of the Roberts dynamo working with a simple fluid flow which is,
with respect to proper Cartesian co--ordinates $x$, $y$ and $z$, 
periodic in $x$ and $y$ and independent of $z$.
A modified Roberts dynamo problem is considered with a flow more similar 
to that in the experimental device.
Solutions are calculated numerically,
and on this basis an estimate of the excitation condition 
of the experimental dynamo is given.
The modified Roberts dynamo problem is also considered in the framework 
of the mean--field dynamo theory, in which the crucial induction effect 
of the fluid motion is an anisotropic $\alpha$--effect.
Numerical results are given for the dependence of the mean--field
coefficients on the fluid flow rates.
The excitation condition of the dynamo is also discussed within this framework.
The behavior of the dynamo in the nonlinear regime, i.e.\ with backreaction
of the magnetic field on the fluid flow, 
depends on the effect of the Lorentz force on the flow rates. 
The quantities determining this effect are calculated numerically.
The results for the mean--field coefficients and the quantities describing the backreaction
provide corrections to earlier results, which were obtained 
under simplifying assumptions.
    
Key words: dynamo, dynamo experiment, mean--field dynamo 
theory, $\alpha$--effect, Lorentz force 
\end{abstract}

\pacs{52.65.Kj, 52.75.Fk, 47.65.+a}

\maketitle

\section{Introduction} 
\label{intro} 
 
In the Forschungszentrum \mbox{Karlsruhe} U.\ M\"uller and R.\ Stieg\-litz
have set up an experimental device
for the demonstration and investigation of a homogeneous dynamo 
as it is expected in the Earth's interior or in cosmic bodies
\cite{stieglitzetal96}. 
The experiment ran first time successfully in December 1999,  
and since then several series of measurements have been carried out  
\cite{muelleretal00,stieglitzetal01,stieglitzetal02,muelleretal02}. 
It is the second realization of a homogeneous dynamo in the
laboratory. Its first run followed only a few weeks after that of the
Riga dynamo experiment, working with a somewhat different principle,
which was pushed forward by A.\ Gailitis, O.\ Lielausis and co--workers
\cite{Gai00,Gai01}.

\begin{figure}\includegraphics[width=.45\textwidth]{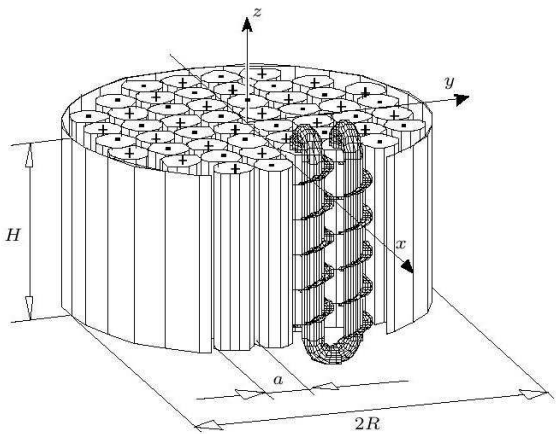}\caption[]{
The dynamo module (after \cite{stieglitzetal96}).
The signs + and -- indicate that the fluid moves in the positive or negative 
$z$--direction, respectively, in a given spin generator.
$R = 0.85\,$m, $H = 0.71\,$m, $a = 0.21\,$m. 
}\label{module}\end{figure}

\begin{figure}\includegraphics[width=.45\textwidth]{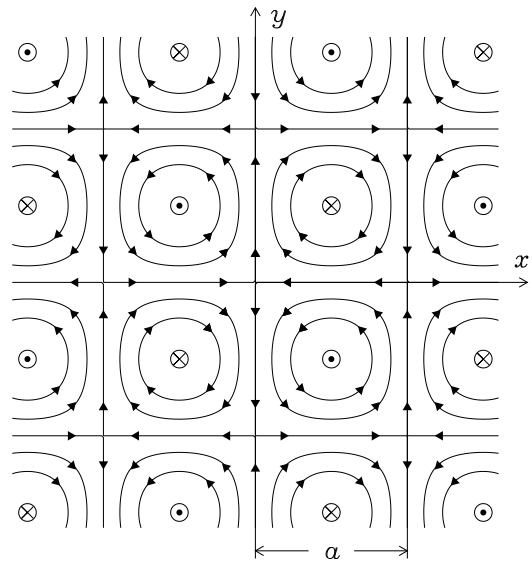}\caption[]{
The Roberts flow pattern. 
The flow directions correspond to the situation in the dynamo module  
if the co--ordinate system coincides with that in Fig.~\ref{module}.
}\label{robflow}\end{figure}

\begin{figure}\includegraphics[width=.45\textwidth]{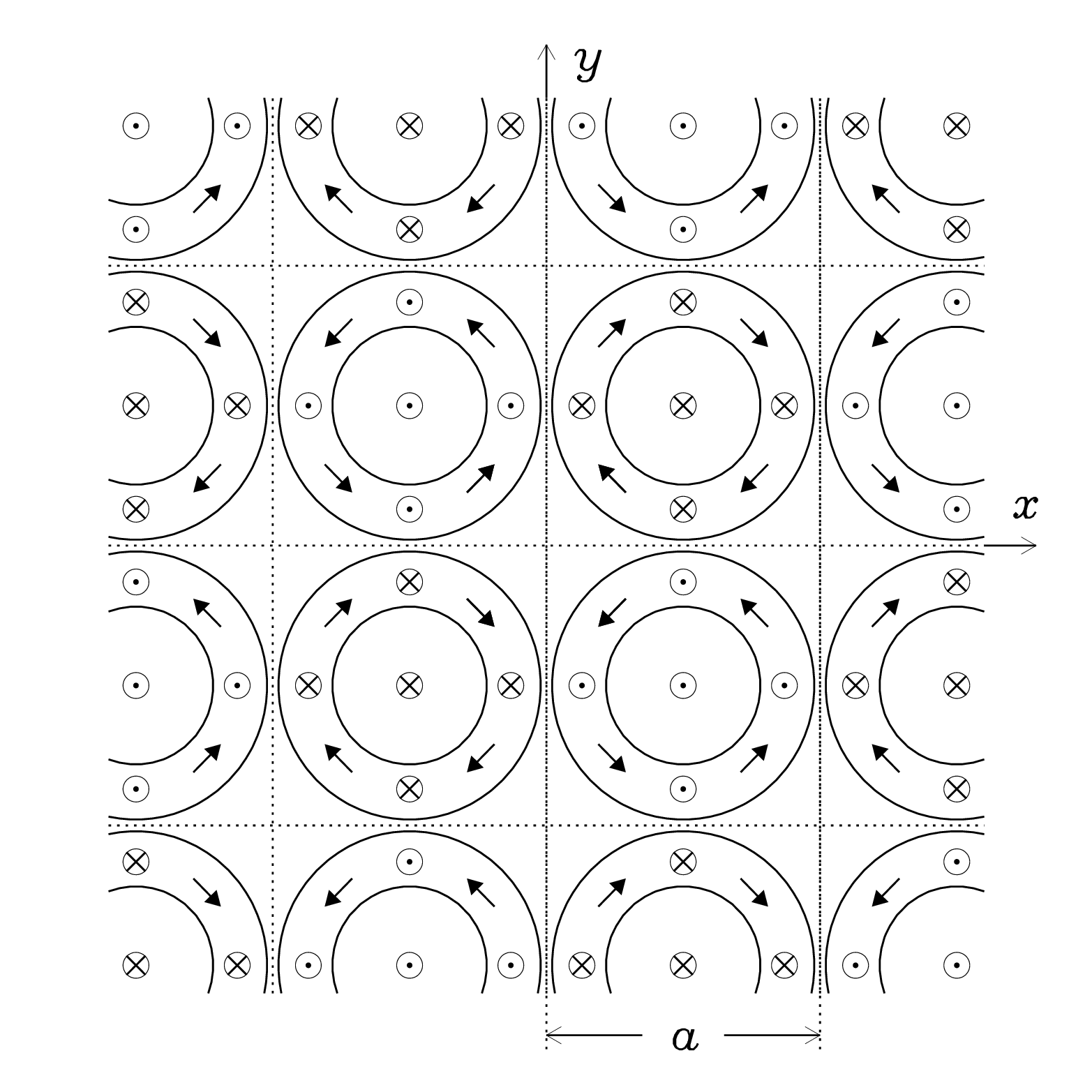}\caption[]{
The spin generator flow pattern.
As for the flow directions the remark given with Fig.~\ref{robflow} applies.
The fluid outside the cylindrical regions where flow
directions are indicated is at rest.
There are no walls between the cells.
}\label{spingenflow}\end{figure}

The basic idea of the Karlsruhe experiment was proposed in 1975 by F.~H.\ Busse
\cite{busse75,busse92}.  
It is very similar to an idea discussed already in 1967 by A.\ Gailitis
\cite{gailitis67}. 
The essential piece of the experimental device, the dynamo module, is a 
cylindrical container  
as shown in Fig.~\ref{module}, with both radius and height somewhat less 
than 1m, through which   
liquid sodium is driven by external pumps. By means of a system of 
channels with conducting  
walls, constituting 52 ``spin generators", helical motions are organized.  
The flow pattern resembles one of those considered in the theoretical work 
of G.~O.\ Roberts in 1972 \cite{robertsgo72}.
This kind of Roberts flow, which proved to be capable of dynamo action, 
is sketched in Fig.~\ref{robflow}. 
In a proper Cartesian co-ordinate system $(x, y, z)$ it is periodic in $x$ and $y$
with the same period length, which we call here $2 a$, but independent of $z$. 
The $x$ and $y$--components of the velocity can be described by a stream function 
proportional to $\sin (\pi x /a) \sin (\pi y /a)$,
and the $z$--component is simply proportional to $\sin (\pi x /a) \sin (\pi y /a)$.
When speaking of a ``cell" of the flow we mean a unit 
like that given by $0 \leq x, y \leq a$.
Clearly the velocity is continuous everywhere, 
and at least the $x$ and $y$--components do not vanish at the margins of the cells.
The real flow in the spin generators deviates from the Roberts flow in the way 
indicated in Fig.~\ref{spingenflow}.
In each cell there are a central channel and a helical channel around it. 
In the simplest approximation the fluid moves rigidly in each of these channels, 
and it is at rest outside the channels.
We relate the word ``spin generator flow" in the following to this simple flow.
In contrast to the Roberts flow the spin generator flow shows discontinuities 
and vanishes at the margins of the cells.       

The theory of the dynamo effect in the Karlsruhe device has been widely elaborated.
Both direct numerical solutions of the induction equation
for the magnetic field \cite{tilgner96,tilgner97,tilgner00,tilgneretal01,
tilgneretal02,tilgner02,tilgner02b}
as well as mean--field theory and solutions of the corresponding equations 
\cite{raedleretal96,raedleretal97a,raedleretal98a,raedleretal02a,raedleretal02b,raedleretal02c}
have been employed.
We focus our attention here on this mean--field approach.
In this context mean fields are understood as averages over areas in planes 
perpendicular to the axis of the dynamo module 
covering the cross--sections of several cells.  
The crucial induction effect of the fluid motion is then, 
with respect to the mean magnetic field, described as an anisotropic $\alpha$--effect.  
The $\alpha$--coefficient and related quantities have first been calculated 
for the Roberts flow 
\cite{raedleretal97a,raedleretal96,raedleretal97b,raedleretal02a,raedleretal02b}. 
In the calculations with the spin generator flow carried out so far,
apart from the case of small flow rates, a simplifying 
but not strictly justified assumption was used.
The contribution of a given spin generator to the $\alpha$--effect was considered 
independent of the neighboring spin generators and in that sense determined 
under the condition that all its surroundings are conducting fluid at rest
\cite{raedleretal97a,raedleretal97b,raedleretal02a,raedleretal02b}.
An analogous assumption was used in calculations of the effect 
of the Lorentz force on the fluid flow rates 
in the channels of the spin generators
\cite{raedleretal02a,raedleretal02c}.
It remained to be clarified which errors result from these assumptions.  

The main purpose of this paper is therefore the calculation of the $\alpha$--coefficient and 
a related coefficient 
as well as the quantities determining the effect of the Lorentz force 
on the fluid flow rates for an array of spin generators,
taking into account the so far ignored mutual influences 
of the spin generators. 
In Section \ref{dynprob} the modified Roberts dynamo problem 
with the spin generator flow is formulated.
In Section \ref{nummeth} the numerical method used for solving 
this problem and the related problems occurring in the following sections
are discussed. 
Section \ref{excond} presents in particular results concerning 
the excitation condition for the dynamo with spin generator flow.
In Section \ref{mfappr} various aspects of a mean--field theory
of the dynamo experiment are explained
and results for the mean electromotive force
due to the spin generator flow are given.
Section \ref{lorentz} deals with the effect of the Lorentz force 
on the flow rates in the channels of the spin generators.
Finally in Section \ref{conclus} some consequences of our findings 
for the understanding of the experimental results are summarized.

Independent of the recent comprehensive accounts of the 
mean--field approach to the Karlsruhe dynamo experiment
\cite{raedleretal02a,raedleretal02b,raedleretal02c},
this paper may serve as an introduction to the basic idea 
of the experiment.
However, we do not strive to repeat all important issues
discussed in those papers, but we mainly want to deliver
the two supplements mentioned above.

\section{Formulation of the dynamo problem}
\label{dynprob}

Let us first formulate the analogue of the Roberts dynamo problem 
for the spin generator flow.
We consider a magnetic field $\bB$ in an infinitely extended homogeneous 
electrically conducting fluid, which is governed by the induction equation, 
\begin{equation} 
\eta \nabla^2 \bB + \nabla \times (\bu \times \bB)
    - \partial_t \bB = {\bf 0} \, , 
    \quad \nabla \cdot \bB = 0 \, , 
\label{eq201} 
\end{equation}
where $\eta$ is the magnetic diffusivity of the fluid
and $\bu$ its velocity. 
The fluid is considered incompressible, so $\nabla \cdot \bu = 0$.
Referring to the Cartesian co--ordinate system $(x, y, z)$ mentioned above 
we focus our attention on the cell $0 \leq x, y \leq a$
and introduce there cylindrical co--ordinates $(r, \varphi, z)$ 
such that the axis $r = 0$ coincides with $x = y = a/2$.
We define then the fluid velocity $\bu$ in this cell by
\begin{eqnarray}
\begin{array}{lll}
u_r = 0&& \quad \mbox{everywhere}\\
u_{\varphi} = 0,& u_z = - u & \quad 
\mbox{for $0< r \le r_1$}\\ 
u_{\varphi} = - \omega r,& u_z = - (h / 2 \pi) \omega & \quad 
\mbox{for $r_1 < r \le r_2$}\\ 
u_{\varphi} = 0,& u_z = 0  & \quad 
\mbox{for $r > r_2$},\\
\end{array}
\label{eq203}
\end{eqnarray} 
where $u$ and $\omega$ are constants, 
$r_1$ and $r_2$ are the radius of the central channel 
and the outer radius of the helical channel, respectively, 
and $h$ is the pitch of the helical channel.
The coupling between $u_{\varphi}$ and $u_z$ in $r_1 < r \le r_2$
considers the constraint on the flow resulting from the helicoidal walls 
of the helical channel.
The velocity $\bu$ in all space follows from the continuation 
of velocity in the considered cell in the way indicated in Fig.~\ref{spingenflow},
i.e.\ with changes of the flow directions from each cell to the adjacent ones 
so that the total pattern is again periodic in $x$ and $y$ 
with the period length $2 a$ and independent of $z$.

We characterize the magnitudes of the fluid flow through the central 
and helical channels of a spin generator 
by the volumetric flow rates $V_{\mathrm{C}}$ and $V_{\mathrm{H}}$ 
given by  
\begin{equation}
V_{\mathrm{C}} = \pi r^2_1 u \, , \quad
V_{\mathrm{H}} = \frac{1}{2} (r^2_2 - r^2_1) h \omega \, .
\label{eq205}
\end{equation} 
We may measure them in units of $a \eta$, so we introduce the dimensionless flow rates  
$\tilde{V}_{\mathrm{C}}$ and $\tilde{V}_{\mathrm{H}}$,
\begin{equation}
\tilde{V}_{\mathrm{C}} = V_{\mathrm{C}} / a \eta \, , \quad
\tilde{V}_{\mathrm{H}} = V_{\mathrm{H}} / a \eta \, .
\label{eq207}
\end{equation} 
We further define magnetic Reynolds numbers $R_{\rm m\mathrm{C}}$ and $R_{\rm m\mathrm{H}}$
for the two channels by $R_{\rm m\mathrm{C}} = u r_1 / \eta$
and $R_{\rm m\mathrm{H}} = \omega r_2 (r_2 - r_1) / \eta$.
Thus we have $\tilde{V}_{\mathrm{C}} = (\pi r_1 / a) R_{\rm m\mathrm{C}}$
and $\tilde{V}_{\mathrm{H}} = [(r_1 + r_2) h / 2 a r_2] R_{\rm m\mathrm{H}}$.
  
In view of the application of the results for the considered dynamo problem  
to the experimental device we mention here the numerical values 
for the radius $R$ and the height $H$ of the dynamo module, 
the lengths $a$, $h$, $r_1$ and $r_2$ characterizing a spin generator
and the magnetic diffusivity $\eta$ of the fluid:
$R = 0.85 \, \mbox{m}$,  
$H = 0.71 \, \mbox{m}$,
$a = 0.21 \, \mbox{m}$,
$h = 0.19 \, \mbox{m}$,
$r_1 / a = 0.25$,
$r_2 / a = 0.5$,
$\eta = 0.1 \, {\mbox{m}}^2 / \mbox{s}$.
(More precisely, the values of $R$ and $H$ apply to the ``homogeneous part"
of the dynamo module, i.e.\ the part without connections 
between different spin generators. 
The value of $\eta$ is slightly higher than that for sodium at $120^\circ\mbox{C}$,
considering the effective reduction of the magnetic diffusivity 
by the steel walls of the channels.)
The given data imply $a \eta = 75.6 \, {\mbox{m}}^3 / \mbox{s}$.
Furthermore we have $\tilde{V}_{\mathrm{C}} = 0.785 R_{\rm m\mathrm{C}}$
and $\tilde{V}_{\mathrm{H}} = 1.357 R_{\rm m\mathrm{H}}$,
so $\tilde{V}_{\mathrm{C}}$ and $\tilde{V}_{\mathrm{H}}$
are in fact magnetic Reynolds numbers. 
Concerning deviations from the rigid--body motion of the fluid 
assumed here and the role of turbulence we refer to the more 
comprehensive representations
\cite{raedleretal02a,raedleretal02b}.

We are interested in dynamo action of the fluid motion, 
so we are interested in growing solutions $\bB$ of (\ref{eq201}) 
with the velocity $\bu$ defined by (\ref{eq203})
and the explanations given with them.
According to some modification of Cowling's anti--dynamo theorem  
growing solutions $\bB$ independent of $z$ are impossible;
cf. \cite{lortz68}.
We restrict our attention to solutions of the form 
\begin{equation}
\bB = \mbox{Re}\big[\hat{\bB} (x, y, t) \exp (\mbox{i}k z)\big] \, ,
\label{eq209}
\end{equation}
where $\hat{\bB}$ is a complex periodic vector field which has again 
a period length $2 a$ in $x$ and $y$, and $k$ a non--vanishing real constant.
In this case we may consider equations (\ref{eq201}) in the period interval
$- a \leq x, y \leq a$ only and adopt periodic boundary conditions.
(Solutions $\bB$ with larger period lengths, as were investigated 
for the Roberts flow \cite{tilgneretal95,plunianetal02a,plunianetal02b},
seem to be well possible but are not considered here.)

If we put $\hat{\bB}(x,y,t) = \hat{\bB}(x,y) \exp(pt)$ with a parameter $p$,
for which we have to admit complex values,
equation (\ref{eq201}) together with the boundary conditions 
pose an eigenvalue problem with $p$ being the eigenvalue parameter.
Clearly $p$ depends on $V_{\mathrm{C}}$, $V_{\mathrm{H}}$ and $k$.
The condition $\mbox{Re}(p) = 0$ defines for each given $k$ 
a neutral line, i.e.\ a line of marginal stability,
in the $V_{\mathrm{C}} V_{\mathrm{H}}$--diagram,
which separates the region of $V_{\mathrm{C}}$ and $V_{\mathrm{H}}$
in which growing $\bB$ are impossible from that where they are possible.

\section{The numerical method}
\label{nummeth}

In view of the numerical solution of the induction equation (\ref{eq201}) 
we express $\bB$ by a vector potential $\bA$,
\begin{equation} 
\bB = \nabla \times \bA \, .
\label{eq301}
\end{equation}
Inserting this in (\ref{eq201}) and choosing $\nabla \cdot \bA$ properly
we may conclude that
\begin{equation}
\eta \nabla^2 \bA + \bu \times \bB - \partial_t \bA = {\bf 0} \, .
\label{eq303} 
\end{equation}
Analogous to (\ref{eq209}) we put 
\begin{equation}
\bA (x,y,z,t) = \mbox{Re}\big[\hat{\bA} (x,y,t) \exp(\mbox{i} k z)\big]\,.
\label{eq305} 
\end{equation}  
Then we have
\begin{equation}
\hat{\bB} =  \nabla \times \hat{\bA} + \mbox{i} \bk \times \hat{\bA},
\label{eq309} 
\end{equation}
where $\bk = k \be$ with $\be$ being the unit vector in $z$--direction, and
\begin{equation}
\eta (\nabla^2 - k^2 )\hat{\bA} + \bu \times \hat{\bB}
    - \partial_t \hat{\bA} = {\bf 0} \,.
\label{eq307} 
\end{equation}
With a solution $\hat{\bA}$ we can calculate $\hat{\bB}$ according to    
({\ref{eq309}) and finally $\bB$ according to (\ref{eq301}).

In the sense explained above we consider (\ref{eq307}) only in the period unit 
$- a \leq x, y \leq a$ and adopt periodic boundary conditions.
When replacing $\hat{\bA}(x,y,t)$ by $\hat{\bA}(x,y) \exp(pt)$ we arrive again 
at an eigenvalue problem with $p$ as eigenvalue parameter.

Let us, for example, assume that $p$ is real and consider the steady case, 
$p = 0$.
We may then consider, e.g., $V_{\mathrm{C}}$ as eigenvalue parameter 
while $V_{\mathrm{H}}$ and $k$ are given.
Modifying the equation resulting from (\ref{eq305}) by an artificial quenching 
of $V_{\mathrm{C}}$ and following up the evolution of $\hat{\bA}$, 
the wanted steady solutions of the original equation (\ref{eq307})
and thus the relations between $V_{\mathrm{C}}$ and $V_{\mathrm{H}}$ 
for given $k$ and $p = 0$ can be found.   
  
For the numerical computations a grid--point scheme was used. 
They were carried out on a two--dimensional mesh typically with 
$60 \times 60$ or $120 \times 120$ points, 
and some of the results were checked with $240 \times 240$ points. 
The $x$ and $y$-derivatives were calculated using sixth order 
explicitly finite differences, 
and the equations were stepped forward in time 
using a third order Runge--Kutta scheme.  

\section{The excitation condition of the dynamo} 
\label{excond}

Using the described numerical method solutions $\bB$ of the dynamo problem 
posed by (\ref{eq201}), (\ref{eq203}) and (\ref{eq209}) have been determined.  
As in the case of the Roberts flow \cite{plunianetal02a,plunianetal02b}
the most easily excitable solutions are non-oscillatory,
which corresponds to real $p$, 
and possess a contribution independent of $x$ and $y$. 

Fig.~\ref{Fptable_crit} shows the neutral lines 
in the $V_{\mathrm{C}} V_{\mathrm{H}}$--diagram      
for several values of the dimensionless quantity $\kappa$ defined by $\kappa = a k$.
In view of the Karlsruhe experiment the case deserves special interest
in which a ``half wave" of $\bB$ fits just to the height $H$ of the dynamo module, 
so $\kappa = \pi a / H = 0.929$.
The neutral line for this case can provide us a very rough estimate 
of the excitation condition of the Karlsruhe dynamo.
However, this estimate neither takes into account the finite radial extend 
of the dynamo module nor realistic conditions at its plane boundaries.
Let us consider, e.g., the values of $V_{\mathrm{H}}$ necessary 
for self--excitation in the experimental device for given $V_{\mathrm{C}}$.
The values of $V_{\mathrm{H}}$ obtained in the experiment 
as well as those found by direct numerical simulations
are by a factor of about 2 higher than the values 
concluded from the neutral curve for $\kappa = 0.9$;
see e.g. Fig.~4 in Refs.~\cite{muelleretal02} 
and \cite{stieglitzetal02}, Fig.~2 in Ref.~\cite{tilgner02} 
or Fig.~3 in Ref.~\cite{tilgner02b}.
The tendency of the variation of $V_{\mathrm{H}}$ with $V_{\mathrm{C}}$ 
is however well predicted.    
(The influence of the finite radial extend of the dynamo module 
on the excitation condition will be discussed in Section \ref{mfexc}.
It makes the mentioned factor of about 2 plausible.) 

\begin{figure}\includegraphics[width=.45\textwidth]{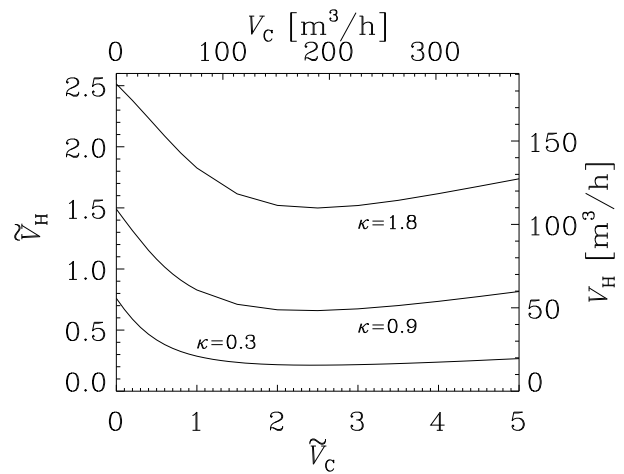}\caption[]{
Neutral lines describing steady dynamo states in the $V_{\rm C} V_{\rm H}$--plane
for various values of $\kappa$.
}\label{Fptable_crit}\end{figure}

\section{The mean--field approach}
\label{mfappr}

The Karlsruhe dynamo experiment has been widely discussed in the framework 
of the mean--field dynamo theory;
see e.g. \cite{krauseetal80}.
Let us first discuss a few aspects of the traditional mean--field approach
applied to spatially periodic flows 
and then a slight modification of this approach, 
which possesses in one respect a higher degree of generality.
We always assume that the magnetic flux density $\bB$ is governed
by the induction equation (\ref{eq201}) and the fluid velocity $\bu$ is specified 
to be either a Roberts flow or the spin generator flow as defined above.

\subsection{The traditional approach}
\label{tradappr}

For each given field $F$ we define a mean field $\overline{F}$
by taking an average over an area corresponding to the cross--section 
of four cells in the $xy$--plane, 
\begin{equation}
\overline{F}(x,y,z)
   = \frac{1}{4a^2} \int \limits_{-a}^a \int \limits_{-a}^a
   F(x+\xi,y+\eta,z) \, d\xi \, d\eta \, .
\label{eq401} 
\end{equation}
We note that the applicability of the Reynolds averaging rules, 
which we use in the following, 
requires that $\overline{F}$ varies only weakly 
over distances $a$ in $x$ or $y$--direction.
(The following applies also with a definition of $\overline{F}$ 
using averages over an area corresponding to two cells only \cite{plunianetal02a}, 
but we do not want to consider this possibility here.)

We split the magnetic flux density $\bB$ and the fluid velocity $\bu$
into mean fields $\overline{\bB}$ and $\overline{\bu}$ 
and remaining fields $\bB'$ and $\bu'$,
\begin{equation}
\bB = \overline{\bB} + \bB' \, , \quad
\bu = \overline{\bu} + \bu' \, .
\label{eq403} 
\end{equation}
Clearly we have $\overline{\bu} = {\bf 0}$, and therefore $\bu = \bu'$. 

Taking the average of equations (\ref{eq201}) we see
that $\overline{\bB}$ has to obey
\begin{equation} 
\eta \nabla^2 \bmB + \nabla \times \bE
   - \partial_t \bmB = {\bf 0} \, ,
   \quad \nabla \cdot \bmB = 0 \, ,
\label{eq405} 
\end{equation}
where $\bE$, defined by
\begin{equation}
\bE = \overline{\bu \times \bB'} \, ,
\label{eq407} 
\end{equation}
is a mean electromotive force due to the fluid motion.

The determination of $\bE$ for a given $\bu$ requires the knowledge of $\bB'$.
Combining equations (\ref{eq201}) and (\ref{eq405}) we easily arrive at  
\begin{eqnarray} 
\eta \nabla^2 \bB' + \nabla \times ( \bu \times \bB')' - \partial_t \bB'  
   &=& - \, \nabla \times (\bu \times \overline{\bB})\,,
\nonumber\\ 
\nabla \cdot \bB' &=& 0 \, , 
\label{eq409} 
\end{eqnarray}
where $(\bu \times \bB')' = \bu \times \bB' - \overline{\bu \times \bB'}$.
We conclude from this that $\bB'$ is, apart from initial and boundary conditions, 
determined by $\bu$ and $\overline{\bB}$ and is linear in $\overline{\bB}$. 
We assume here that $\bB'$ vanishes if $\overline{\bB}$ does so
(and will comment on this below). 
Thus $\bE$ too can be understood as a quantity determined by $\bu$ and 
$\overline{\bB}$ only and being linear and homogeneous in $\overline{\bB}$.
Of course, $\bE$ at a given point in space and time depends not simply on $\bu$ 
and $\overline{\bB}$ in this point but also on their behavior in some neighborhood 
of this point.  

We adopt the assumption often used in mean--field dynamo theory 
that $\overline{\bB}$ varies only weakly in space and time 
so that $\overline{\bB}$ and its first spatial derivatives in this point are sufficient 
to define the behavior of $\overline{\bB}$ in the relevant neighborhood. 
Then $\bE$ can be represented in the form
\begin{equation} 
{\cal{E}}_i = a_{ij} \, \overline{B}_j  
 + b_{ijk} \, \partial \overline{B}_j / \partial x_k \, ,  
\label{eq411} 
\end{equation}  
where the tensors $a_{ij}$ and $b_{ijk}$ are averaged quantities determined by $\bu$. 
We use here and in the following the notation $x_1 = x$, $x_2 = y$, $x_3 = z$ 
and adopt the summation convention. 
Of course, the neglect of contributions to $\bE$ with higher order spatial derivatives 
or with time derivatives of $\overline{\bB}$ 
(which is in one respect relaxed in Section \ref{modappr}) 
remains to be checked in all applications.

The specific properties of the considered flow patterns 
allow us to reduce the form of $\bE$ given by (\ref{eq411}) 
to a more specific one.  
Due to our definition of averages and the periodicity of the flow patterns
in $x$ and $y$, and its independence of $z$,
the tensors $a_{ij}$ and $b_{ijk}$ are independent of $x$, $y$ and $z$.
Clearly a $90^\circ$ rotation of the flow pattern about the $z$--axis
as well as a shift by a length $a$ along the $x$ or $y$--axes 
change only the sign of $\bu$ so that simultaneous rotation and shift 
leave $\bu$ unchanged. 
This is sufficient to conclude that $a_{ij}$ and $b_{ijk}$ 
are axisymmetric tensors with respect to the $z$--axis.
So $a_{ij}$ and $b_{ijk}$ contain no other tensorial construction elements
than the Kronecker tensor $\delta_{lm}$,
the Levi--Civita tensor $\epsilon_{lmn}$
and the unit vector $\be$ in $z$--direction.
The independence of the flow pattern of $z$ requires that
$a_{ij}$ and $b_{ijk}$ are invariant under the change of the sign of $\be$.
Finally it can be concluded on the basis of (\ref{eq409}) that $\bE$ has to vanish
if $\bmB$ is a homogeneous field in $z$--direction, which leads to $a_{33} = 0$.     
With the specification of $a_{ij}$ and $b_{ijk}$ according to these requirements 
relation (\ref{eq411}) turns into
\begin{eqnarray}
\bE
 & = & - \alpha_{\perp} \big(\overline{\bB} - (\be \cdot \overline{\bB}) \, \be \big) 
\label{eq413}\nonumber \\
 & & - \beta_{\perp} \nabla \times \overline{\bB}  
 - (\beta_{\parallel}-\beta_{\perp}) (\be \cdot (\nabla \times \overline{\bB})) \, \be
\nonumber\\  
 & & - \beta_3 \be \times (\nabla (\be \cdot \overline{\bB}) 
 + (\be \cdot \nabla) \overline{\bB}) \, , 
\end{eqnarray} 
where the coefficients $\alpha_{\perp}$, $\beta_{\perp}$, $\beta_{\parallel}$
and $\beta_3$ are averaged quantities determined by $\bu$ 
and independent of $x$, $y$ and $z$.
The term with $\alpha_{\perp}$ describes an $\alpha$--effect,
which is extremely anisotropic. 
It is able to drive electric currents in the $x$ and $y$--directions,
but not in the $z$--direction.
The terms with $\beta_{\perp}$ and $\beta_{\parallel}$ give rise to the introduction 
of a mean-field diffusivity different from the original magnetic diffusivity 
of the fluid and again anisotropic. 
In contrast to them the remaining term with $\beta_3$ is not connected 
with $\nabla \times \overline{\bB}$ but with the symmetric part 
of the gradient tensor of $\overline{\bB}$
and can therefore not be interpreted in the sense 
of a mean-field diffusivity.

In the case of the Roberts flow the coefficient $\alpha_\perp$ has been determined 
for arbitrary flow rates, 
and coefficients like $\beta_\perp$, $\beta_\parallel$ and $\beta_3$ 
for small flow rates
\cite{raedleretal96,raedleretal97a,raedleretal02a,raedleretal97b,raedleretal02b}.
As for the spin generator flow only results for $\alpha_\perp$ have been given so far
\cite{raedleretal96,raedleretal97a,raedleretal97b,raedleretal02b,raedleretal02a}.

For the determination of $\alpha_\perp$ it is sufficient 
to consider equation (\ref{eq409}) for $\bB'$ with $\bmB$ 
specified to be homogeneous field. 
In this case, which implies $\nabla \times (\overline{\bu \times \bB'}) = {\bf 0}$,
this equation turns into 
\begin{eqnarray}
\eta \nabla^2 \bB' + (\bB' \cdot \nabla) \bu - (\bu \cdot \nabla) \bB'
    - \partial_t \bB' &=& - \, (\bmB \cdot \nabla) \bu\,,
\nonumber\\
    \nabla \cdot \bB' &=& 0 \, .
\label{eq415}
\end{eqnarray} 
We may again consider $\bB'$ like $\bmB$ as independent of $z$.
Let us put $\bB' = {\bB'}_\perp + {\bB'}_\parallel$ 
and $\bu = \bu_\perp + \bu_\parallel$ 
with ${\bB'}_\perp = \bB' - (\be \cdot \bB') \be$ 
and ${\bB'}_\parallel = (\be \cdot \bB') \be$, 
and $\bu_\perp$ and $\bu_\parallel$ defined analogously. 
Then we find 
\begin{eqnarray}
\eta \nabla^2 {\bB'}_\perp 
   &+& ({\bB'}_\perp \cdot \nabla) \bu_\perp 
   - (\bu_\perp \cdot \nabla) {\bB'}_\perp
   - \partial_t {\bB'}_\perp
\nonumber\\
   && \quad \quad \quad = - (\bmB \cdot \nabla) \bu_\perp,
\nonumber\\
\eta \nabla^2 {\bB'}_\parallel
   &-& (\bu_\perp \cdot \nabla) {\bB'}_\parallel
   - \partial_t {\bB'}_\parallel
\label{eq417}\\
   && \quad \quad \quad = - ((\bmB + {\bB'}_\perp) \cdot \nabla) \bu_\parallel \, .
\nonumber
\end{eqnarray}
We further put $\bu_\perp = u_\perp {\tilde{\bu}}_\perp$ 
and $\bu_\parallel = u_\parallel {\tilde{\bu}}_\parallel$,
where $u_\perp$ and $u_\parallel$ are factors independent of $x$ and $y$
characterizing the magnitudes of $\bu_\perp$ and $\bu_\parallel$,
and ${\tilde{\bu}}_\perp$ and ${\tilde{\bu}}_\parallel$ fields 
which are normalized in some way.   
Clearly ${\bB'}_\perp$ is independent of $u_\parallel$,
and ${\bB'}_\parallel$ linear in $u_\parallel$.
The $x$ and $y$--components of $\overline{\bu \times \bB'}$,
from which $\alpha_\perp$ can be concluded,
are sums of products of components of $\bu_\parallel$ and ${\bB'}_\perp$
and of $\bu_\perp$ and ${\bB'}_\parallel$.
Thus $\alpha_\perp$ must depend in a homogeneous and linear way on $u_\parallel$
whereas the dependence on $u_\perp$ is in general more complex.
This can be observed from the results for the Roberts flow. 
In view of the spin generator flow we split $\bu_\parallel$ into two parts,
$\bu_{\parallel \, 1}$ and $\bu_{\parallel \, 2}$,
of which the first one is non--zero in the central channel and the second one in the 
helical channel only.
We further introduce the corresponding quantities $u_{\parallel \, 1}$ 
and $u_{\parallel \, 2}$.
We may then conclude that $\alpha_\perp$ is linear 
but no longer homogeneous in $u_{\parallel \, 1}$.
Since $u_{\parallel \, 1}$ is proportional to $V_{\mathrm{C}}$
we find that $\alpha_\perp$ is linear but not homogeneous in $V_{\mathrm{C}}$
whereas it shows a more complex dependence on $V_{\mathrm{H}}$. 

For small flow rates we may neglect the terms with $\bu$ on the left--hand side 
of equation (\ref{eq415}).
This corresponds to the second--order correlation approximation often 
used in mean--field dynamo theory.
Then the solutions $\bB'$ and further $\alpha_\perp$ can be calculated analytically.
Starting from the result found in this way for the spin generator flow 
\cite{raedleretal96,raedleretal97a,raedleretal97b} and using the above findings
we conclude that the general form of $\alpha_\perp$ reads
\begin{equation}
\alpha_\perp = \frac{V_{\mathrm{H}}}{a^2 h \eta}
   \left[ V_{\mathrm{C}} \phi_{\mathrm{C}} (V_{\mathrm{H}} / h \eta)
   + {\textstyle\frac{1}{2}}V_{\mathrm{H}} \phi_{\mathrm{H}} (V_{\mathrm{H}} / h \eta) \right] 
\label{eq419}
\end{equation} 
with two functions $\phi_{\mathrm{C}}$ and $\phi_{\mathrm{H}}$ 
satisfying $\phi_{\mathrm{C}} (0) = \phi_{\mathrm{H}} (0) = 1$.
Note that the argument $V_{\mathrm{H}} / h \eta$ is equal to
$(a / h) {\tilde{V}}_{\mathrm{H}}$, 
which is in turn equal to  
$\omega (r_2 + r_1) (r_2 - r_1) / 2 \eta$. 
Consequently it is just some kind of magnetic Reynolds number 
for the rotational motion of the fluid in a helical channel.  
The functions $\phi_{\mathrm{C}}$ and $\phi_{\mathrm{H}}$ 
have been calculated analytically under a simplifying assumption
\cite{raedleretal97a,raedleretal97b}, 
which, however, proved not to be strictly correct.
We will give rigorous results for $\alpha_\perp$ 
and for $\phi_{\mathrm{C}}$ and $\phi_{\mathrm{H}}$
in a more general context below in Section \ref{alphaetc}.

As announced we make now a comment on the assumption that $\bB'$ vanishes 
if $\bmB$ does so. 
Investigations with the Roberts dynamo problem have revealed 
that non--decaying solutions $\bB$ of the induction equation (\ref{eq201}) 
whose average over a cell vanishes are well possible \cite{plunianetal02a}.
They coincide with non--decaying solutions $\bB'$ of the equation (\ref{eq409})
in the case $\bmB = {\bf 0}$.
These solutions are, however, always less easily excitable than solutions 
with non--vanishing averages over a cell. 
They are therefore without interest in the discussion of the excitation condition
for mean magnetic fields $\bmB$. 
In that sense the above assumption is, although not generally true, 
at least in the case of the Roberts flow acceptable for our purposes.
Presumably this applies also for the spin generator flow. 

In view of the next Section we assume for a moment that $\overline{\bB}$ 
does not depend on $x$ and $y$ but only on $z$.
In that case we have $\nabla \times \overline{\bB} =   
\be \times [\nabla (\be \cdot \overline{\bB}) + (\be \cdot \nabla) \overline{\bB}]
= \be \times \mbox{d} \overline{\bB} / \mbox{d} z$ 
and therefore (\ref{eq413}) turns into
\begin{eqnarray}
\bE = - \alpha_{\perp} \big(\overline{\bB} - (\be \cdot \overline{\bB}) \, \be \big)
   - \beta \be \times \mbox{d} \overline{\bB} / \mbox{d} z\,,\nonumber \\
\beta = \beta_\perp + \beta_3 \, .
\label{eq420}
\end{eqnarray}
Interestingly enough, here the difference in the characters 
of the $\beta_\perp$ and $\beta_3$--terms in (\ref{eq413}) 
is no longer visible.
While there are reasons to assume that the coefficients 
$\beta_\perp$ and $\beta_\parallel$, which can be interpreted 
in the sense of a mean--field diffusivity, are never negative, 
this is no longer true for $\beta_3$ 
and therefore also not for $\beta$.
The results for the Roberts flow show indeed explicitly that $\beta$
can take also negative values 
\cite{raedleretal96,raedleretal02a,raedleretal02b}.

\subsection{A modified approach}
\label{modappr}

We now modify the mean--field approach discussed so far in view of the case 
in which $\overline{\bB}$ does not depend on $x$ and $y$ but may have 
an arbitrary dependence on $z$.
All quantities like $\bB$, $\overline{\bB}$, $\bB'$ or $\bE$, 
which depend on $z$, are represented as Fourier integrals according to  
\begin{equation}
F (x,y,z,t) = \int \hat{F} (x,y,k,t) \exp(\mbox{i} k z) \mbox{d} k \, .  
\label{eq421} 
\end{equation}
The corresponding representation of $\bB$ clearly includes the ansatz (\ref{eq209}).
$\hat{\bB}$ depends on $x$, $y$, $k$ and $t$, 
but $\hat{\overline{\bB}}$ and $\hat{\bE}$ depend only on $k$ and $t$.
The requirement that $F (x,y,z,t)$ is real leads to  
$F^* (x,y,k,t) = F (x,y,-k,t)$.
Relations of this kind apply to $\bB$, $\bmB$, $\bB'$ and $\bE$.

Equations (\ref{eq401}) to (\ref{eq409}) remain valid
whereas (\ref{eq411}), (\ref{eq413}) and (\ref{eq420}) have to be modified.
Clearly (\ref{eq405}) and (\ref{eq407}) are equivalent to
\begin{equation} 
\eta k^2 \hat{\bmB} - \mbox{i} \bk \times \hat{\bE}
   + \partial_t \hat{\bmB} = {\bf 0} \, ,
   \quad \be \cdot \hat{\bmB}  = 0 \, ,
\label{eq423} 
\end{equation}
and
\begin{equation}
\hat{\bE} = \overline{\bu \times \hat{\bB}'}.
\label{eq425} 
\end{equation}
Instead of (\ref{eq409}) we have  
\begin{eqnarray} 
\eta (\nabla^2 - k^2) \hat{\bB'} 
   + (\nabla + \mbox{i} \bk) \times (\bu \times \hat{\bB}')'
   - \partial_t \hat{\bB}' 
\nonumber\\
= - (\nabla + \mbox{i} \bk) \times (\bu \times \hat{\bmB})\,,
\quad (\nabla + \mbox{i} \bk) \cdot \hat{\bB}' = 0 \, , 
\label{eq427}
\end{eqnarray}
where $(\bu \times \hat{\bB}')' = \bu \times \hat{\bB}' - \overline{\bu \times \hat{\bB}'}$.

Assuming again that $\bE$ is linear and homogeneous in $\bmB$ we conclude 
that the same applies to $\hat{\bE}$ and $\hat{\bmB}$, too.
Therefore we now have
\begin{equation}
{\hat{\cal{E}}}_i (k, t)  
    =  {\hat{\alpha}}_{ij} (k) {\hat{\overline{B}}}_j (k, t) \, ,
\label{eq429} 
\end{equation}
where ${\hat{\alpha}}_{ij}$ is a complex tensor determined by the fluid flow. 
Analogous to $\hat{\bE}$ and $\hat{\bmB}$ it has to satisfy 
${\hat{\alpha}}_{ij}^* (k) = {\hat{\alpha}}_{ij} (-k)$. 
From the symmetry properties of the $\bu$--field we conclude again 
that the connection between $\bE$ and $\bmB$ remains its form if both 
are simultaneously subject to a $90^\circ$ rotation about the $z$--axis, 
i.e.\ relation (\ref{eq429}) remains unchanged under such a rotation 
of $\hat{\bE}$ and $\hat{\bmB}$.
This means that the tensor ${\hat{\alpha}}_{ij}$ is axisymmetric 
with respect to the axis defined by $\bk$.
The general form of ${\hat{\alpha}}_{ij}$ that is
compatible with ${\hat{\alpha}}_{ij}^* (k) = {\hat{\alpha}}_{ij} (-k)$
is given by 
\begin{equation}
{\hat{\alpha}}_{ij} (k) 
    = a_1 (|k|) \delta_{ij} + a_2 (|k|) k_i k_j 
    + \mbox{i} a_3 (|k|) \epsilon_{ijl} k_l 
\label{eq431} 
\end{equation}
with real $a _1$, $a_2$ and $a_3$.
Together with (\ref{eq429}) this leads to 
${\hat{\cal{E}}}_z = (a_1 + a_2 k^2) {\hat{\overline{B}}}_z$. 
On the other hand ${\hat{\cal{E}}}_z$ is equal to the average of
$u_x {\hat{B}'}_y - u_y {\hat{B}'}_x$, 
and we may conclude from (\ref{eq427}) that ${\hat{B}'}_x$ and ${\hat{B}'}_y$
are independent of ${\hat{\overline{B}}}_z$.
This in turn implies $a_1 + a_2 k^2 = 0$.
We note the final result for ${\hat{\alpha}}_{ij} (k)$ in the form
\begin{equation}
{\hat{a}}_{ij} (k) = - {\hat{\alpha}}_\perp (k) (\delta_{ij} - e_i e_j) 
   + \mbox{i} {\hat{\beta}} (k) \epsilon_{ijl} k_l 
\label{eq433}
\end{equation}
with two real quantities ${\hat{\alpha}}_\perp$ and ${\hat{\beta}}$, 
which are even functions of $k$. 

From (\ref{eq429}) and (\ref{eq433}) we obtain
\begin{equation}
\hat{\bE} (k) = - {\hat{\alpha}}_\perp (k)\big(\hat{\bmB} - (\be \cdot \hat{\bmB}) \be\big)
   -\mbox{i} \hat{\beta} (k) \bk \times \hat{\bmB} \, .
\label{eq435}
\end{equation}
Together with (\ref{eq421}) this leads to 
\begin{eqnarray}
&&\bE (z, t) = 
\nonumber\\ 
&&\quad - \int {\hat{\alpha}}_\perp (k) \big[ \hat{\bmB} (k, t) 
   - (\be \cdot \hat{\bmB} (k, t)) \be \big] 
   \exp(\mbox{i}kz) \mbox{d}k 
\nonumber\\
&&\quad - \, \be \times \frac{\partial}{\partial z} 
    \int \hat{\beta} (k) \hat{\bmB} (k, t) 
    \exp(\mbox{i}kz) \mbox{d}k \, . 
\label{eq437}
\end{eqnarray}
This in turn is equivalent to
\begin{eqnarray}
&&\bE (z, t) = 
\nonumber\\
&&\quad - \frac{1}{2 \pi} 
    \int \alpha_\perp (\zeta) \big[ \hat{\bmB} (z + \zeta, t) 
    - (\be \cdot \hat{\bmB} (z + \zeta, t)) \be \big] 
    \mbox{d}\zeta 
\nonumber \\
&&\quad - \frac{1}{2 \pi} \be \times \frac{\partial}{\partial z} 
    \int \beta (\zeta) \hat{\bmB} (z + \zeta, t) 
    \mbox{d} \zeta 
\label{eq439}
\end{eqnarray}
with
\begin{eqnarray}
\alpha_\perp (\zeta) 
   &=& \int {\hat{\alpha}}_\perp (k) \exp( \mbox{i} k \zeta ) \mbox{d} k\,,
\nonumber\\
\beta (\zeta) 
   &=& \int {\hat{\beta}} (k) \exp( \mbox{i} k \zeta ) \mbox{d} k \, .
\label{eq441} 
\end{eqnarray}
Note that both $\alpha_\perp$ and $\beta$ are even in $\zeta$.

Let us now expand $\hat{\alpha}_{ij} (k)$ as given by (\ref{eq433}) in a Taylor series
and truncate it after the second term,
\begin{equation}
{\hat{\alpha}}_{ij} (k) 
    = - {\hat{\alpha}}_\perp (0) ( \delta_{ij} - e_i e_j )  
    + \mbox{i} k \hat{\beta} (0) \epsilon_{ijl} e_l  \, .
\label{eq443}
\end{equation} 
The corresponding expansion of $\bE$ as given by (\ref{eq437}) reads
\begin{equation}
\bE = - {\hat{\alpha}}_\perp (0)  \big(\bmB - (\be \cdot \bmB) \be \big)
     - \hat{\beta} (0)\, \be \times \mbox{d} \bmB / \mbox{d} z  \, .
\label{eq445}
\end{equation}
Comparing this with relation (\ref{eq420}) of the preceding section 
we find
\begin{equation} 
\alpha_\perp =  {\hat{\alpha}}_\perp (0) \, , \quad 
    \beta = \hat{\beta} (0) \, .
\label{eq447}
\end{equation} 

Returning again to arbitrary $k$ we define for later purposes 
a function $\hat{\alpha} (k)$ by
\begin{equation} 
\hat{\alpha} (k) =  {\hat{\alpha}}_\perp (k) + k \hat{\beta} (k) \, .
\label{eq449}
\end{equation} 
If $\hat{\alpha} (k)$ is given, we may determine 
${\hat{\alpha}}_\perp (k)$ and $\hat{\beta} (k)$ according to
\begin{eqnarray}
\hat{\alpha}_\perp (k) &=& \frac{1}{2}\big[\hat{\alpha}(k) + \hat{\alpha}(-k)\big]\,,
\nonumber\\
\beta (k) &=& \frac{1}{2 k}\big[\hat{\alpha}(k) - \hat{\alpha}(-k)\big]  \, .
\label{eq451}
\end{eqnarray}
Moreover, we have
\begin{equation} 
\alpha_\perp  =  \hat{\alpha}(0)\, , \quad 
    \beta =  \frac{\mbox{d} \hat{\alpha}(k)}{\mbox{d} k} (0) \, .
\label{eq453}
\end{equation}

\subsection{The parameters defining $\alpha$--effect etc.}
\label{alphaetc}

In view of the determination of the quantities 
$\hat{\alpha}_\perp (k)$ and $\hat{\beta} (k)$,
which includes that of $\alpha_\perp$ and $\beta$,
we note that relations like (\ref{eq420}) or (\ref{eq435}) 
connecting $\bE$ with $\bmB$ or $\hat{\bE}$ with $\hat{\bmB}$ apply, 
apart from the explicitly mentioned restrictions, for arbitrary $\bmB$.
Thus we may take these quantities from calculations carried out for specific $\bmB$.

Using the method described in Section \ref{nummeth} we have numerically determined 
steady solutions of equation (\ref{eq409}) for $\hat{\bB}'$ 
with given $V_{\mathrm{C}}$, $V_{\mathrm{H}}$, $k$ 
and a specific $\hat{\bmB}$ of Beltrami type satisfying 
$\be \times \mbox{d} \hat{\bmB} / \mbox{d} z = k \hat{\bmB}$.
With these solutions we have then calculated the quantity 
$\hat{\bE} \cdot {\hat{\bmB}}^*$, which, according to (\ref{eq435}), has to satisfy 
\begin{equation} 
\hat{\bE} \cdot {\hat{\bmB}}^* = - \hat{\alpha} \big|\hat{\bmB}\big|^2
\label{eq471}
\end{equation}
with $\hat{\alpha}$ defined by (\ref{eq449}).
From the values of $\hat{\alpha}$ and their dependence on $k$ 
obtained in this way $\hat{\alpha}_\perp (k)$, $\hat{\beta} (k)$, 
$\alpha_\perp$ and $\beta$ have been determined.

In mean--field models of the Karlsruhe device in the sense 
of the traditional approach explained in Section \ref{tradappr}
the coefficient $\alpha_\perp$ occurs in the dimensionless combination 
$C = \alpha_\perp R / \eta$, with $R$ being the radius of the dynamo module,
and the influence of $\beta$ can be discussed in terms of
$\tilde{\beta} = \beta / \eta$.
We generalize the definitions of $C$ and $\tilde{\beta}$ by putting
\begin{equation} 
C = {\hat{\alpha}}_\perp R / \eta \, , \quad
    \tilde{\beta} = \hat{\beta} / \eta \, .
\label{eq473}
\end{equation}
Now $C$ and $\tilde{\beta}$ show a dependence on $k$, which we express
by one on $\kappa = a k$.

Thinking first of the traditional approach we consider $C$ with $\kappa = 0$.
Figure~\ref{Fptable_axes} shows contours of $C$ in  
the $V_{\mathrm{C}}V_{\mathrm{H}}$--diagram, 
Fig.~\ref{fig8} the functions $\phi_{\mathrm{C}}$ and $\phi_{\mathrm{H}}$,
from which $\alpha_\perp$ and thus $C$ can be calculated.
These results deviate for large $V_{\mathrm{H}}$ significantly from those determined
with the simplifying assumption mentioned above, according to which 
the mutual influence of the spin generators was ignored
\cite{raedleretal97a,raedleretal97b}.
In the region of $V_{\mathrm{C}}$ and $V_{\mathrm{H}}$ 
which is of interest for the experiment,
say $0 < {\tilde{V}}_{\mathrm{C}}, {\tilde{V}}_{\mathrm{H}} < 2$, 
the values of $C$, $\phi_{\mathrm{C}}$ and $\phi_{\mathrm{H}}$ 
for given $V_{\mathrm{C}}$ and $V_{\mathrm{H}}$
are somewhat larger than those obtained with that assumption.
One reason for that might be that in the case of an array of spin generators, 
compared to a single one in a fluid at rest,
the rotational motion in a helical channel expels less magnetic flux 
into regions without fluid motion, where it can not contribute to the $\alpha$-effect.
Remarkably, in the region $0 < {\tilde{V}}_{\mathrm{C}}, {\tilde{V}}_{\mathrm{H}} < 2$
our result for $C$ agrees very well with one derived under the assumption 
of a Roberts flow \cite{raedleretal02a,raedleretal02b}. 

Figure~\ref{fig9} exhibits contours of $\tilde{\beta}$ for
$\kappa=0$ in the $V_{\mathrm{C}}V_{\mathrm{H}}$--diagram.
We already pointed out that $\tilde{\beta}$ can take negative values.
Here we see that $\tilde{\beta}$ becomes negative for sufficiently 
large values of $V_{\mathrm{C}}$ and $V_{\mathrm{H}}$.
Although this happens somewhat beyond the region of interest for the experiment
it suggests that inside this region the positive values of $\tilde{\beta}$ 
may be small.
The diffusion term in the mean--field induction equation is proportional 
to $\eta (1 + \tilde{\beta})$. 
In the investigated region of $V_{\mathrm{C}}$ and $V_{\mathrm{H}}$
this quantity proved always to be positive.

Let us now proceed to $C$ and $\tilde{\beta}$ for $\kappa \not= 0$.
As already mentioned, in view of the experimental device it seems reasonable 
to put $\kappa = \pi a / H = 0.929$.
Analogous to Fig.~\ref{Fptable_axes}, which applies to $\kappa = 0$,
Fig.~\ref{Fptable_diff} shows contours of $C$ for $\kappa = 0.9$.
We see that $C$ for given $V_{\mathrm{C}}$ and $V_{\mathrm{H}}$ 
is slightly higher in the latter case.
The results for $\tilde{\beta}$ are virtually indistinguishable for both cases.

\begin{figure}\includegraphics[width=.45\textwidth]{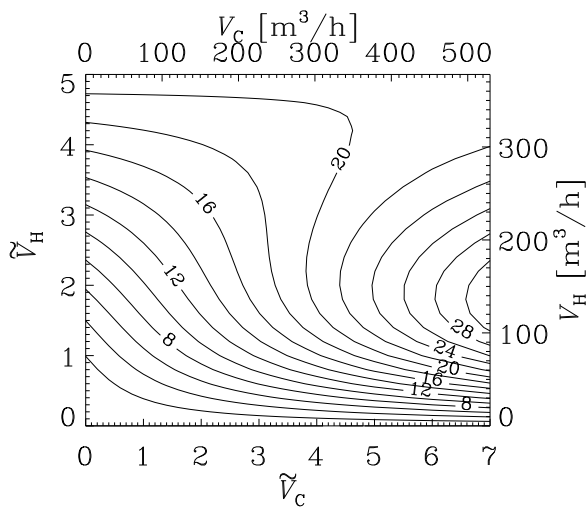}\caption[]{
Contours of $C$ for $\kappa = 0$.
}\label{Fptable_axes}\end{figure}

\begin{figure}\includegraphics[width=.45\textwidth]{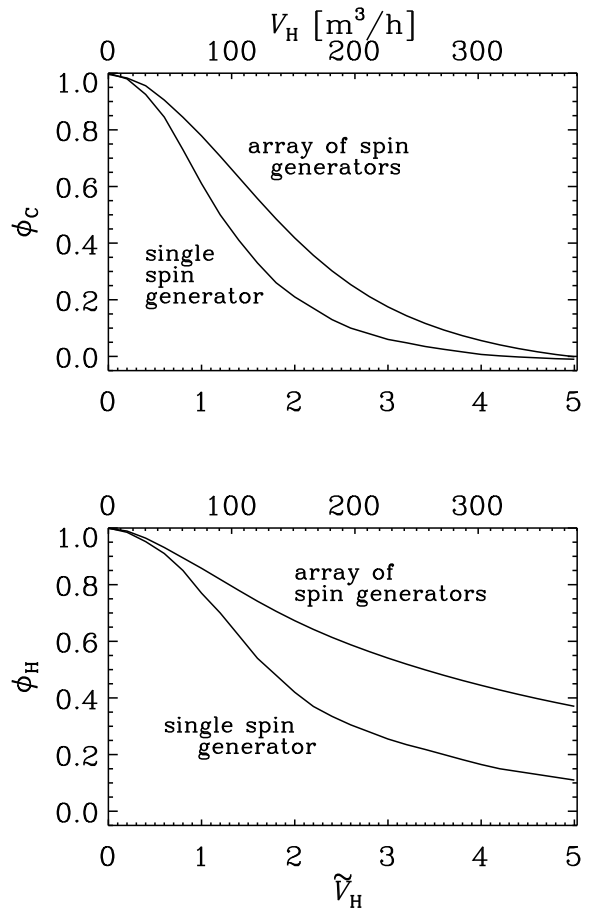}\caption[]{
The functions $\phi_{\rm{C}}$ and $\phi_{\rm{H}}$ calculated for an array 
of spin generators. For comparison the results of the approximation 
considering single spin generators (i.e. ignoring their mutual influences)
are also given.
}\label{fig8}\end{figure}

\begin{figure}\includegraphics[width=.45\textwidth]{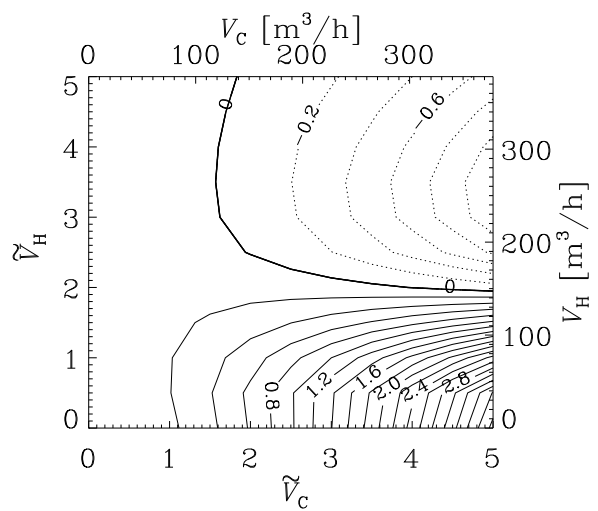}\caption[]{
Contours of $\tilde{\beta}$ for $\kappa=0$
(calculated as the limit $\kappa \to 0$).
}\label{fig9}\end{figure}

\begin{figure}\includegraphics[width=.45\textwidth]{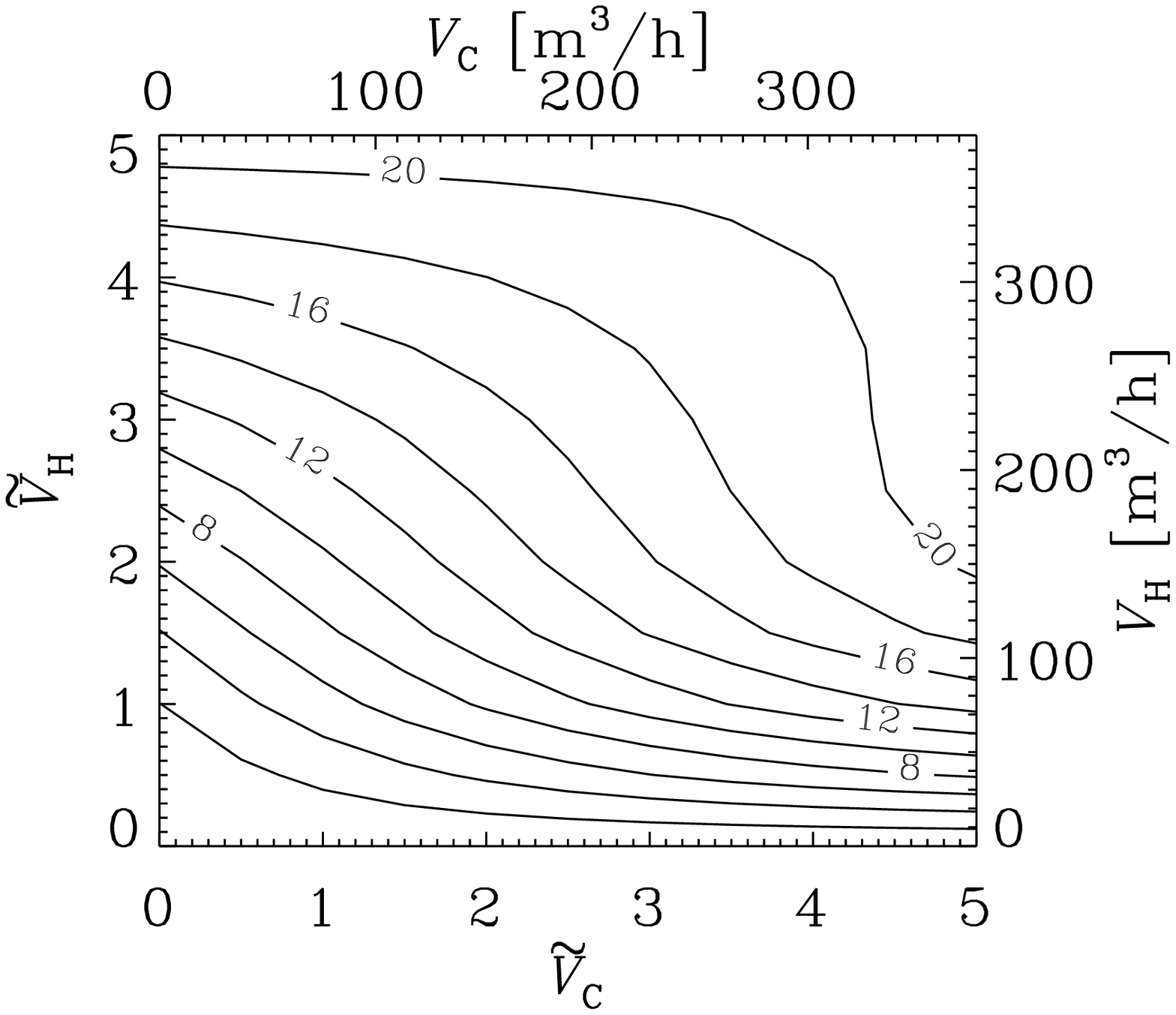}\caption[]{
Contours of $C$ for $\kappa = 0.9$.
}\label{Fptable_diff}\end{figure}

\subsection{The excitation condition in mean--field models}
\label{mfexc}

We consider first again the traditional approach to mean--field theory explained 
in Section \ref{tradappr}. 
Equation (\ref{eq405}) for $\bmB$ together with relation (\ref{eq413}) for $\bE$
allows the solutions 
\begin{eqnarray} 
\bmB &=& B_0 \big(\cos (kz), \mp \sin (kz), 0\big) \exp (pt)\,,
\nonumber\\
p &=& - (\eta + \beta) k^2 \pm \alpha_\perp k \, ,
\label{eq481}
\end{eqnarray}
where $B_0$ is an arbitrary constant.
We refer here again to Cartesian co--ordinates 
and consider $k$ as a positive parameter.
For these solutions we have $\nabla \times \bmB = \pm k \bmB$, i.e.\
they are of Beltrami type.
This implies that there are no mean electric currents in the $z$--direction.   
The solution that corresponds to the upper signs can grow if $\alpha_\perp$
is sufficiently large.
The condition of marginal stability reads $\alpha_\perp = (\eta +\beta) k$ or, 
what is the same,
\begin{equation}
C = (1 + \tilde{\beta}) k R \, , 
\label{eq483}
\end{equation}
where $C$ and $\tilde{\beta}$ have to be interpreted as the values for $\kappa = 0$.
If we relate this to the dynamo module and put $k = \pi / H$ we have
\begin{equation}
C = (1 + \tilde{\beta}) \pi R / H \, . 
\label{eq485}
\end{equation}
Note that the factor $R$ in the conditions (\ref{eq483}) and (\ref{eq485})
results from the definition of $C$ only.   
In fact they are independent of $R$.

Proceeding to the modified approach to the mean--field theory 
and replacing relation (\ref{eq413}) for $\bE$ by (\ref{eq437}) 
we find formally the same result.
However, $\alpha_\perp$ and $\beta$ have to be replaced 
by $\hat{\alpha}_\perp$ and $\hat{\beta}$, 
and $C$ and $\tilde{\beta}$ in (\ref{eq483}) and (\ref{eq485}) 
have to be taken for $\kappa = a k$.
The condition (\ref{eq483}) interpreted in this sense 
defines neutral lines in the $V_{\mathrm{C}} V_{\mathrm{H}}$--diagram
which have to agree exactly with those shown in Fig.~\ref{Fptable_crit}. 
Likewise, the condition (\ref{eq485}) defines the special neutral line with $\kappa = \pi a/H$.

One of the shortcomings of estimates 
of the self--excitation condition of the experimental device
based on the solutions of the induction equation used in Section \ref{excond} 
or, equivalently, on a relation like (\ref{eq485}),
consists in ignoring the finite radial extent of the dynamo module. 
We point out another solution of equation (\ref{eq405}) for $\bmB$, 
which has been used for an estimate of the self-excitation condition 
of the experimental device considering its finite radial extent
\cite{busseetal98,raedleretal97a,raedleretal02b}.
For the sake of simplicity we assume that $\bE$ is given 
by equation (\ref{eq413}) with $\beta_\perp = \beta_\parallel = \beta_3 = 0$.
We refer to a  new cylindrical co--ordinate system $(r, \varphi, z)$ 
adjusted to the dynamo module so that $r = 0$ coincides with its axis 
and $z = 0$ with its midplane.
The solution we have in mind reads
\begin{eqnarray} 
\bmB &\!=\!& B_0 \left( \, \frac{\partial {\it \Psi}}{\partial z}, \, 
   \frac{\eta (q^2 + k^2) + p}{\alpha_\perp} {\it \Psi}, \,
   - \frac{1}{r} \frac{\partial}{\partial r} (r {\it \Psi}) \right)
   \exp (pt)\,,
\nonumber\\
{\it \Psi} &\!=\!& J_0 (q r) \cos (kz)\,,
\label{eq487}\\
p &\!=\!& - \eta (q^2 + k^2) \pm \alpha_\perp k  \, ,
\nonumber
\end{eqnarray}
where $q$ and $k$ are constants and $J_0$ is the zero--order Bessel function 
of the first kind.      
This solution is axi\-symmetric with respect to the $z$--axis.
It has further the property that the normal components of  $\nabla \times \bmB$ 
vanish both on the cylindrical surfaces $q r = z_\nu$, 
where $z_\nu$ denotes the zeros of $J_0$,
and on the planes $kz = (l + 1/2) \pi$ with integer $l$.
We identify the region inside the smallest of these cylindrical surfaces 
between two neighboring planes of that kind with the dynamo module, 
so we put $q = z_1 / R$, where $z_1$ is the smallest positive zero of $J_0$, 
and $k =\pi / H$.  
Then there are no electric currents penetrating the surface of the dynamo module.
The condition of marginal stability for the so specified solution reads
\begin{equation}
C =  \pi (R / H) \big[ 1 + (z_1 H / \pi R)^2 \big] \, .
\label{eq489}
\end{equation}
In the limit $H/R \to 0$ this agrees with (\ref{eq485}) 
if we put $\tilde{\beta} = 0$.
For finite $H/R$, however, $C$ is now always larger than the value 
given by (\ref{eq485}) with $\tilde{\beta} = 0$. 
This can easily be understood considering that there is now 
an additional dissipation of the magnetic field due to its radial gradient.
$C$ as function of $H/R$ has a minimum at $H/R = \pi/z_1$.
The dynamo module was designed so that $H/R$ has just this value. 
In this case we have 
\begin{equation}
C =  2 \pi R /H  \, . 
\label{eq491}
\end{equation}
In other words, the real radial extent of the dynamo module enlarges 
the requirements for $C$, compared to the case of infinite extent, 
by a factor 2.     
As can be seen from Fig.~\ref{Fptable_axes}, 
in the region of $V_{\mathrm{C}}$ and $V_{\mathrm{H}}$ in which
experimental investigations have been carried out,
say $1.3 < {\tilde{V}}_{\mathrm{C}}, {\tilde{V}}_{\mathrm{H}} < 1.6$,
this enlargement of $C$ means that if, e.g., $V_{\mathrm{C}}$ is given,
$V_{\mathrm{H}}$ grows by a factor between 2.5 and 3.5.
We recall here the deviation of the experimental results
from the estimate of the self--excitation condition
given in Section \ref{excond} on the basis of Fig.~\ref{Fptable_crit},
which just corresponds to (\ref{eq485}).
In the light of these explanations concerning the influence of the radial extent
of the dynamo module this deviation is quite plausible.
It is actually rather small, which indicates that our reasoning despite a number 
of neglected effects does not underestimate the requirements 
for self--excitation.

We note that also the result (\ref{eq491}) is not a completely satisfying estimate 
of the self--excitation condition of the experimental device. 
Apart from the fact that it does not consider realistic boundary conditions 
for the dynamo module it is based on an axisymmetric solution of the equation for $\bmB$.
Several investigations have however revealed that a non--axisymmetric solution 
is slightly easier to excite than axisymmetric ones
\cite{raedleretal96,raedleretal98a,raedleretal02a,raedleretal02b}.
The influence of the $\beta_\perp$ and $\beta_3$--terms of $\bE$
can no longer be expressed by $\tilde{\beta}$, 
and there is also an influence of the $\beta_\parallel$--term.
All these influences increase the marginal values of $C$ \cite{raedleretal96}.

\section{The effect of the Lorentz force on the flow rates} 
\label{lorentz}

In the theory of the experiment, equations determining the fluid flow rates 
in the loops containing the central channels and in those containing helical channels
have been derived from the balance of the kinetic energy in these loops.
The rate of change of the kinetic energy in a loop is given by the work done 
by the pumps against the hydraulic resistance and the Lorentz force.
For the work done by the Lorentz force averaged over a central 
or a helical channel we write $\langle \bu \cdot \bff \rangle {\cal V}$,
where $\langle \cdots \rangle$ means the average over this channel, 
${\cal V}$ its volume and $\bff$ the Lorentz force per unit volume,
\begin{equation}
\bff = \mu^{-1} (\nabla \times \bB) \times \bB \, ,
\label{eq501}
\end{equation}
with $\mu$ being the magnetic permeability of free space.

We use again $\bB = \bmB + \bB'$. For all results reported here we have assumed 
that $\bmB$ is a homogeneous field and, correspondingly, 
$\bB'$ is also independent of $z$ so that equations (\ref{eq415}) apply.
Then also $\bff$ is independent of $z$ and $\langle \cdots \rangle$ 
may simply be interpreted as an average over the section of the channel 
with the $xy$--plane.
     
We have calculated the quantities $\langle \bu \cdot \bff \rangle_{\mathrm{C}}$ 
and $\langle \bu \cdot \bff \rangle_{\mathrm{H}}$
for a central and a helical channel analytically in two different approximations
\cite{raedleretal02a,raedleretal02c}.
In approximation (i) all contributions to $\bff$ of higher than first order 
in $V_{\mathrm{C}}$ or $V_{\mathrm{H}}$ were neglected so that it applies 
to small $V_{\mathrm{C}}$ and $V_{\mathrm{H}}$ only.
In approximation (ii) arbitrary $V_{\mathrm{C}}$ and $V_{\mathrm{H}}$ were admitted,
but as in an earlier calculation of the $\alpha$--effect 
only a single spin generator surrounded by conducting medium at rest was considered,
i.e.\ any influence of the neighboring spin generators was ignored.
We represent the results of both approximations in the form
\begin{eqnarray}
\langle \bu \cdot \bff \rangle_{\mathrm{C}}
   &=& - \frac{\sigma}{2 \gamma_{\mathrm{C}}} 
   \left( \frac{V_{\mathrm{C}}}{s_{\mathrm{C}}} \right)^2 
   B^2_\perp \psi_{\mathrm{C}} ( V_{\mathrm{C}}, V_{\mathrm{H}})\,,
\nonumber\\
\langle \bu \cdot \bff \rangle_{\mathrm{H}}
   &=& - \frac{\sigma}{2 \gamma_{\mathrm{H}}} 
   \left( \frac{V_{\mathrm{H}}}{s_{\mathrm{H}}} \right)^2 
   B^2_\perp \psi_{\mathrm{H}} ( V_{\mathrm{C}}, V_{\mathrm{H}}) \, .
\label{eq503}
\end{eqnarray}
Here $\sigma$ is the electric conductivity of the fluid, 
$\gamma_{\mathrm{C}}$ and $\gamma_{\mathrm{H}}$ are given by 
\begin{equation}
\gamma_{\mathrm{C}} = 1 \, , \quad
\gamma_{\mathrm{H}}
   = \frac{(r_1 + r_2)^2 + (h / \pi)^2}{2(r_1^2 + r_2^2) + (h / \pi)^2} \, ,
\label{eq505}
\end{equation}
$s_{\mathrm{C}}$ and $s_{\mathrm{H}}$ are the cross--sections 
of the central and helical channels,
and $B_\perp$ is the magnetic flux density perpendicular to the axis
of the spin generator, i.e.\ to the $z$--axis.
In approximation (i) we have 
$\psi_{\mathrm{C}} = \psi_{\mathrm{H}} = 1$.
In approximation (ii) $\psi_{\mathrm{C}}$ and $\psi_{\mathrm{H}}$ 
are functions of $V_{\mathrm{C}}$ and $V_{\mathrm{H}}$ 
satisfying $\psi_{\mathrm{C}} (V_{\mathrm{C}}, 0) = 1$ for $V_{\mathrm{C}} \not= 0$
and $\psi_{\mathrm{H}} (V_{\mathrm{C}}, 0) = 1$ for  all $V_{\mathrm{C}}$, 
varying only slightly with $V_{\mathrm{C}}$ and decaying with growing $V_{\mathrm{H}}$;
see also Fig.~\ref{Fppsih}.
The factors $\psi_{\mathrm{C}}$ and $\psi_{\mathrm{H}}$ in the relations (\ref{eq503})
for $\langle \bu \cdot \bff \rangle_{\mathrm{C}}$
and $\langle \bu \cdot \bff \rangle_{\mathrm{H}}$ 
describe the reduction of the Lorentz force by the magnetic flux expulsion 
out of the moving fluid by its azimuthal motion. 

We may conclude from the relevant equations 
that \mbox{$\langle \bu \cdot \bff \rangle_{\mathrm{C}}$}
and $\langle \bu \cdot \bff \rangle_{\mathrm{H}}$
can again be represented in the form (\ref{eq503}) if the complete array 
of spin generators and arbitrary $V_{\mathrm{C}}$ and $V_{\mathrm{H}}$ 
are taken into account. 
Only the dependences of $\psi_{\mathrm{C}}$ and $\psi_{\mathrm{H}}$ 
on $V_{\mathrm{C}}$ and $V_{\mathrm{H}}$ changes.

Before giving detailed results we make a general statement on these dependences.
As in the considerations in the paragraph containing (\ref{eq415})
we may again introduce the quantities 
${\bB'}_\perp$, ${\bB'}_\parallel$, $\bu_\perp$, $\bu_\parallel$ 
and use (\ref{eq417}).
With the same reasoning as applied there we find that for the spin generator flow
${\bB'}_\perp$ is independent of $V_{\mathrm{C}}$ 
and ${\bB'}_\parallel$ linear in $V_{\mathrm{C}}$.
We further express $\bff_\perp$ and $\bff_\parallel$, defined analogous 
to ${\bB'}_\perp$ and ${\bB'}_\parallel$,
according to (\ref{eq501}) 
by the components of ${\bB'}_\perp$ and ${\bB'}_\parallel$,
their derivatives and the components of $\bmB$.
In this way we find that $\langle \bu \cdot \bff \rangle_{\mathrm{C}}$
is a sum of two terms, one proportional to $V_{\mathrm{C}}$ 
and the other proportional to $V^2_{\mathrm{C}}$.
Consequently, $\psi_{\mathrm{C}}$ has the form 
$\psi^{(0)}_{\mathrm{C}} (V_{\mathrm{H}}) 
+ \psi^{(-1)}_{\mathrm{C}} (V_{\mathrm{H}}) V^{-1}_{\mathrm{C}}$ 
with $\psi^{(0)}_{\mathrm{C}} (0) = 1$.   
We further find that $\langle \bu \cdot \bff \rangle_{\mathrm{H}}$
is a sum of three terms, one independent of $V_{\mathrm{C}}$ 
and the others proportional to $V_{\mathrm{C}}$ and $V^2_{\mathrm{C}}$, 
and $\psi_{\mathrm{H}}$ has the form 
$\psi^{(0)}_{\mathrm{H}} (V_{\mathrm{H}}) 
+ \psi^{(1)}_{\mathrm{H}} (V_{\mathrm{H}}) V_{\mathrm{C}}
+ \psi^{(2)}_{\mathrm{H}} (V_{\mathrm{H}}) V^2_{\mathrm{C}}$ 
with $\psi^{(0)}_{\mathrm{H}} (0) = 1$.   
This can be seen explicitly from the calculations in the approximation (ii)
mentioned above, in which, by the way,  $\psi^{(2)}_{\mathrm{H}} = 0$.

We have calculated $\psi_{\mathrm{C}}$ and $\psi_{\mathrm{H}}$ numerically 
on the basis of equations (\ref{eq409}) 
using the method described in Section \ref{nummeth}.
The result is shown in Fig.~\ref{Fppsih}.
Instead of the complete array of spin generators we have also considered 
an array in which fluid motion occurs only in one out of $4\times4$ spin generators.
The numerical result obtained for this case agrees very well 
with the analytical result of approximation (ii) shown in Fig.~\ref{Fppsih}.

For a complete array of spin generators the factors $\psi_{\mathrm{C}}$ and 
$\psi_{\mathrm{H}}$ in the relations (\ref{eq503}) are generally larger 
compared to approximation (ii).
In other words, the Lorentz force is less strongly reduced by the azimuthal motion of the fluid.
This can be understood by considering that less magnetic flux can be pushed 
into regions without fluid motion.

\begin{figure}\includegraphics[width=.45\textwidth]{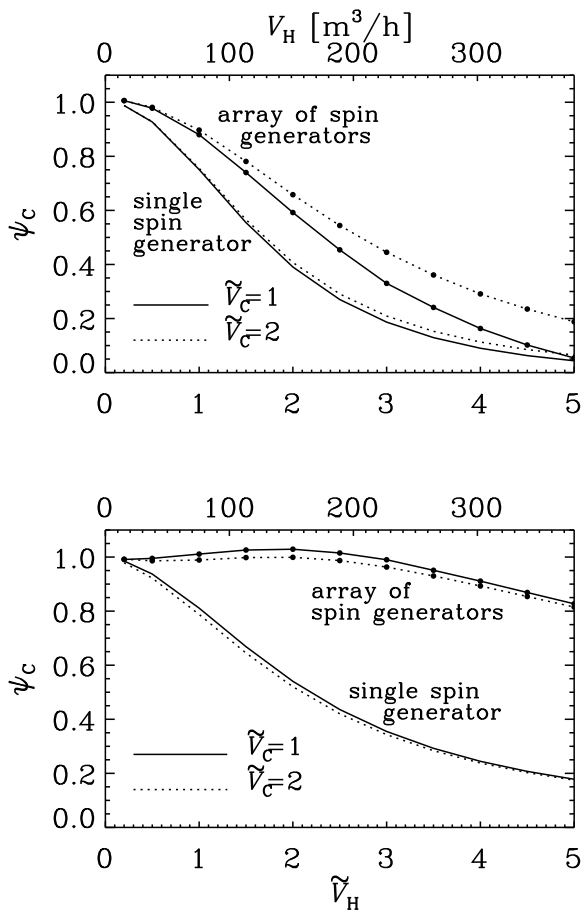}\caption[]{
The dependence of $\psi_{\rm H}$ and $\psi_{\rm C}$ 
on $\tilde{V}_{\rm H}$ for $\tilde{V}_{\rm C} = 1$ and $\tilde{V}_{\rm C} = 2$
for an array of spin generators.
For comparison the results of approximation (ii), which considers a single
spin generator only, are also given.
}\label{Fppsih}\end{figure}

\section{Conclusions}
\label{conclus}

We have first dealt with a modified Roberts dynamo problem with a flow pattern
resembling that in the Karls\-ruhe dynamo module.
Based on numerical solutions of this problem a self--excitation condition 
was found.
Since in these calculations neither the finite radial extent of the dynamo module 
nor realistic boundary conditions at its plane boundaries were taken into account
this self--excitation condition deviates markedly from that 
for the experimental device.

A mean--field approach to the modified Roberts dynamo problem is presented. 
Two slightly different treatments are considered, 
assuming as usual only weak variations 
of the mean magnetic field in space, 
or admitting arbitrary variations in the $z$--direction.
The coefficient $\alpha_\perp$ describing the $\alpha$--effect 
and a coefficient $\beta$ connected with derivatives of the mean magnetic field  
are calculated for arbitrary fluid flow rates.
The result for $\alpha_\perp$ corrects earlier results obtained 
in an approximation that ignores the mutual influences of the spin generators
\cite{raedleretal97b}.
It leads to a much better agreement of the calculated self--excitation condition
with the experimental results
\cite{raedleretal02a,raedleretal02b}.
We note in passing that in the case of small flow rates our result, 
although calculated for rigid--body motions only, 
applies also for more general flow profiles 
\cite{raedleretal02a,raedleretal02b}.   
The result for $\beta$ suggests that the enlargement 
of the effective magnetic diffusivity by the fluid motion 
can be partially compensated by another effect of this motion.
The same has been observed in investigations with the Roberts flow
\cite{plunianetal02b}.
This could be one of the reasons why the results calculated 
under idealizing assumptions, 
in particular ignoring the effect of the mean--field diffusivity, 
deviate only little from the experimental results
\cite{raedleretal02b}.

In the framework of the mean--field approach we have also given an estimate 
of the excitation condition which considers the finite radial extent
of the dynamo module.
It shows that the real extent enhances the critical value of $C$, 
which is a dimensionless measure of $\alpha_\perp$, by a factor 2.
In other words, if in the region of $V_{\mathrm{C}}$ and $V_{\mathrm{H}}$, in
which experimental investigations have been carried out, $V_{\mathrm{C}}$ is fixed,
$V_{\mathrm{H}}$ has to be larger by a factor between 2.5 and 3.5.
If the excitation condition is corrected in this way
it does not underestimate the requirements for self--excitation.

We have also calculated the effect of the Lorentz force on the fluid flow rates 
in the channels of a spin generator.
Again our result corrects a former one obtained in the approximation already mentioned 
which ignores the mutual influences of the spin generators
\cite{raedleretal02a,raedleretal02b}.
The braking effect of the Lorentz force proves to be stronger than predicted 
by the former calculations. 
This means in particular that estimates of the saturation field strengths 
given so far \cite{raedleretal02a,raedleretal02c} 
have to be corrected by factors between 0.8 and 0.9;
for more details see Note added in proof in \cite{raedleretal02c}.

\bigskip

{\bf Acknowledgement}
The results reported in this paper have been obtained during stays of K.-H.R. 
at NORDITA. He is grateful for its hospitality. An anonymous referee is
acknowledged for making useful suggestions.

\end{document}